\documentclass[aps,pra,showpacs,twocolumn,superscriptaddress,nolongbibliography,floatfix]{revtex4-2}
\usepackage{graphicx}
\usepackage{amsmath}
\usepackage{xparse}
\usepackage{dcolumn}
\usepackage{bm}
\usepackage[colorlinks=true, citecolor=blue,allcolors=blue]{hyperref}
\usepackage{physics} 
\usepackage{comment}

\begin{document}
\title{Resonant photoionization dynamics during optical trapping of lithium atoms
}

\author{ K.~Foster}
\affiliation{Physics Department and LAMOR, Missouri University of Science \& Technology, Rolla, MO 65409, USA}

\author{ S.~Majumdar}
\affiliation{Physics Department and LAMOR, Missouri University of Science \& Technology, Rolla, MO 65409, USA}

\author{ D.~Fischer}
\affiliation{Physics Department and LAMOR, Missouri University of Science \& Technology, Rolla, MO 65409, USA}

\date{\today}
\begin{abstract}
Photoionization induced by trapping and auxiliary laser fields is an inherent feature of many laser-cooling and optical trapping experiments, yet its microscopic dynamics are rarely investigated directly. In this work, we employ a reaction microscope implementing an event-by-event photoionization time retrieval technique to extend momentum-resolved photoelectron spectroscopy to continuous-wave laser--atom interactions. We investigate low-intensity multicolor photoionization of laser-cooled lithium atoms confined in an all-optical trap. Complete three-dimensional electron momentum distributions and kinetic-energy spectra recorded for different laser wavelengths and polarization configurations identify resonant excitation of the $5p$ and $5f$ states and reveal an additional ionization channel following spontaneous decay from the $5f$ to the $4d$ state. A pronounced polarization dependence of the photoionization yield is explained by magnetic-sublevel selection rules and the coherent interference of different virtual excitation pathways. These results demonstrate that reaction microscopy combined with event-by-event time reconstruction provides a powerful approach for investigating microscopic electronic dynamics in laser-cooled atomic systems and offers new insight into photoionization processes occurring during optical trapping.
\end{abstract}


\maketitle

\section{Introduction}

Optical dipole traps have become an indispensable tool in modern atomic, molecular, and optical physics since the pioneering work on optical trapping by Ashkin \cite{Ashkin1970} and the subsequent development of far-off-resonant optical dipole traps \cite{Grimm2000}. Their ability to confine laser-cooled atoms with minimal perturbation has enabled applications ranging from precision spectroscopy and ultracold collision studies to quantum simulation and quantum information processing \cite{Ludlow2015,Chin2010,Bloch2008,Gross2017,Saffman2010}. During the loading and operation of these traps, atoms are routinely exposed to several optical fields—including cooling, trapping, repumping, and auxiliary excitation lasers—that can drive unwanted excitation and photoionization. While photoionization induced by trapping and auxiliary laser fields is well recognized as a source of atom loss and reduced trap lifetime \cite{Dinneen1992,Markert2010,Witkowski2022}, it is typically characterized through macroscopic observables such as trap lifetime, atom number, or total ion yield \cite{Wang2023,Hou2025}. Consequently, the underlying electronic excitation pathways and photoionization dynamics responsible for these losses remain comparatively unexplored.

Direct insight into these microscopic processes can be obtained through photoelectron spectroscopy, which provides detailed information about the electronic dynamics underlying atomic photoionization \cite{Starace2023}. The kinetic energy of the emitted electrons identifies the intermediate and final electronic states involved in the ionization process, while the photoelectron angular distribution reflects the symmetry of the populated states and the interference between different ionization pathways \cite{Manson1982}. Momentum imaging techniques such as cold-target recoil-ion momentum spectroscopy (COLTRIMS) provide access to complete three-dimensional electron momentum distributions, enabling detailed investigations of resonant excitation pathways, continuum partial-wave composition, and quantum interference in atomic photoionization \cite{Doerner2000,Ullrich2003}. While these methods have been widely employed to investigate strong-field ionization \cite{Krausz2009} and ultrafast light–matter interactions \cite{Calegari2016}, comparatively few studies have applied momentum-resolved photoelectron spectroscopy to the low-intensity, continuous-wave optical fields routinely used for laser cooling and optical trapping. This is particularly true for laser-cooled alkali atoms, where photoionization is typically monitored through its effect on trap performance rather than through direct measurements of the emitted photoelectrons.

In the present work, we investigate low-intensity multicolor photoionization of laser-cooled lithium atoms confined in an all-optical trap using a reaction microscope \cite{Thini2020, Romans2025a} implementing a recently developed event-by-event photoionization time retrieval technique \cite{Romans2025}, thereby extending momentum-resolved photoelectron spectroscopy to continuous-wave excitation. The combination of continuous-wave multicolor excitation and momentum-resolved electron detection provides complete three-dimensional electron momentum distributions together with electron kinetic-energy spectra, allowing resonant excitation pathways and competing ionization channels to be identified. The measured photoelectron angular distributions provide direct insight into the symmetry of the populated electronic states and the role of quantum interference in the ionization process. Beyond elucidating the microscopic mechanisms responsible for photoionization in optical trapping experiments, this work demonstrates that reaction microscopy can be successfully extended to continuous-wave laser--atom interactions, providing a powerful new approach for investigating electronic dynamics in laser-cooled atomic systems.

\section{Experimental Methods}

The experiment combines an all-optical trap (AOT) for the preparation of laser-cooled and polarized lithium atoms, a continuous-wave two-color excitation scheme for resonance-enhanced multiphoton ionization, and a COLTRIMS reaction microscope for the momentum-resolved detection of emitted electrons and recoil ions. While the AOT and reaction microscope have been described in detail elsewhere \cite{Sharma2018,Hubele2015,Romans2025}, the experimental configuration relevant to the present photoionization measurements is summarized below. 

\subsection{All-optical trap and target preparation}
\label{sec:aot}
The experiment was performed using an AOT for laser-cooled $^6$Li atoms integrated into a COLTRIMS reaction microscope. Similar to conventional magneto-optical traps (MOTs), the AOT employs Doppler cooling provided by three pairs of mutually orthogonal, counterpropagating near-resonant laser beams. Unlike a conventional MOT, the AOT does not require magnetic-field gradients, allowing it to operate in the homogeneous magnetic field needed for momentum-resolved electron spectroscopy. A detailed description and characterization of the trapping scheme have been reported previously \cite{Sharma2018}. Under the present operating conditions, the trap contains approximately $10^7$ atoms at temperatures of a few millikelvin and peak number densities of about $10^9\,\mathrm{cm^{-3}}$. The cooling and repumping beams remain on throughout the measurements, but their broad profiles result in intensities of only a few tens of mW/cm$^2$, substantially lower than those of the tightly focused lasers used for the subsequent excitation and ionization. Approximately 20\,\% of the trapped atoms occupy the excited $2\,^2P_{3/2}$ state. Owing to optical pumping in the presence of the homogeneous magnetic field, this excited-state population is strongly spin-polarized, with the majority of atoms prepared in a single magnetic sublevel, thereby providing a well-defined initial state for the photoionization measurements.

The homogeneous magnetic field defines the quantization axis throughout the experiment and serves as the reference axis for the polarization-dependent measurements discussed below.

\subsection{Excitation and ionization laser configuration}
\label{sec:lasers}

Photoionization was induced by the combination of a high-power continuous-wave infrared laser (hereafter referred to as the trapping laser) and a weak continuous-wave optical laser (hereafter referred to as the coupling laser). The trapping laser was provided by a ytterbium fiber laser operating at 1070\,nm. It was focused to a waist of approximately $50\,\mu$m at the center of the atomic cloud, corresponding to a peak intensity of approximately $1\,\mathrm{MW/cm^2}$. The infrared laser simultaneously provides an optical dipole trap for the lithium atoms and participates directly in the photoionization process by supplying one photon for a resonant two-photon excitation and a second photon for the subsequent ionization step.
According to the manufacturer's specifications, the infrared laser has a spectral bandwidth of $1.5$--$3\,\mathrm{nm}$ (FWHM). Consequently, narrow resonance features are not expected to be resolved, and the laser wavelengths do not require particularly precise adjustment during the measurements.

The coupling laser was provided by a tunable continuous-wave external-cavity diode laser operating either at 677.1\,nm or 673.3\,nm, depending on the excitation scheme under investigation. It was originally implemented for photoassociation experiments on ultracold Li$_2$ molecules, but in the present work serves as the coupling laser for the resonant two-color excitation \cite{Kurz2021}. The beam was spatially overlapped with the infrared trapping beam and focused to a comparable waist at the interaction region. Its optical power was approximately $4\,\mathrm{mW}$, corresponding to a peak intensity almost four orders of magnitude lower than that of the trapping laser. The spatial overlap of the two beams was optimized by maximizing the photoelectron signal.

\begin{figure}
    \centering
    \includegraphics[width=\linewidth]{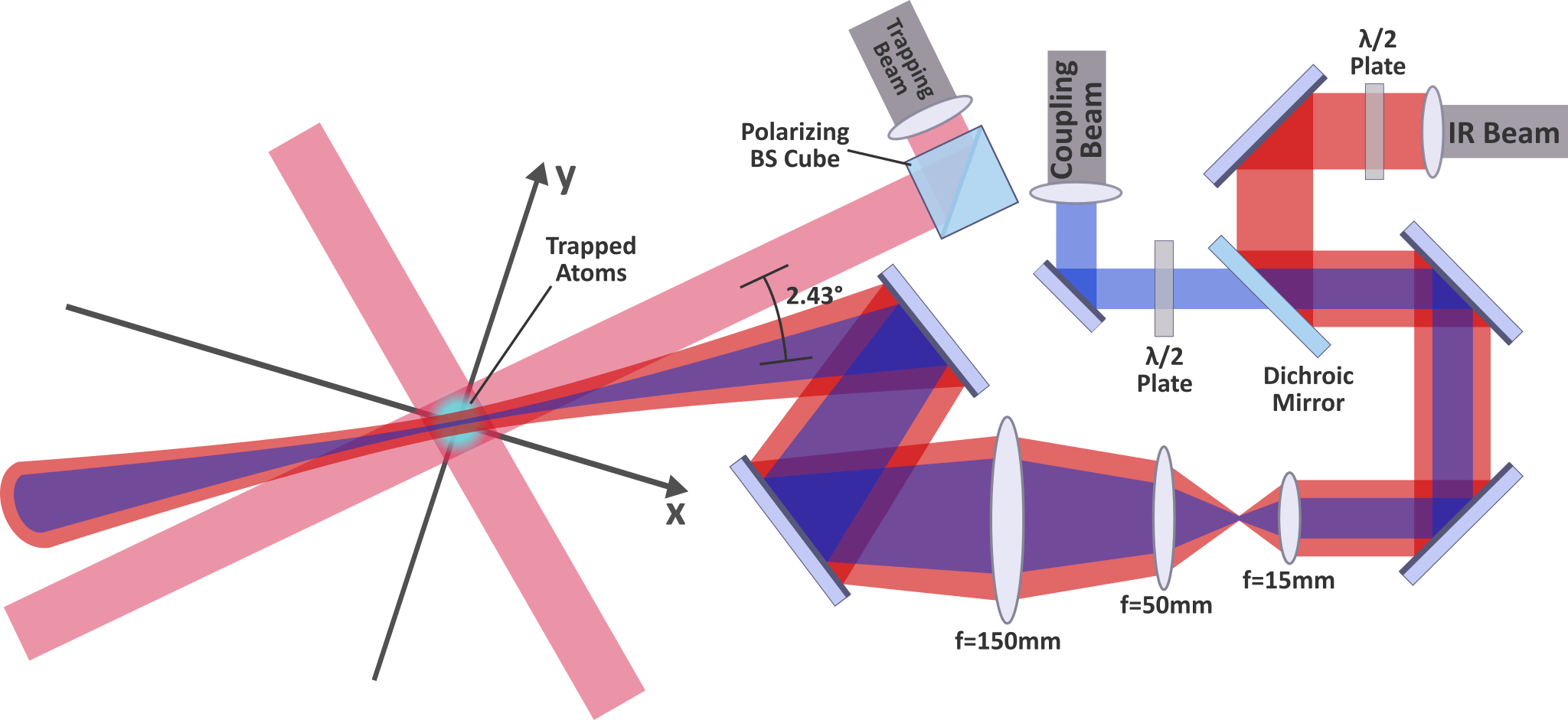}
    \caption{\label{fig:exprDiagram}Diagram of the laser configuration viewed along the negative $z$ direction toward the trapped atoms. In this view, the electron detector is located behind the atomic cloud along the positive $z$ axis, while the ion detector is located in front of the cloud along the negative $z$ axis.}
\end{figure}

The propagation geometry of the laser beams is illustrated in Fig.~\ref{fig:exprDiagram}. Both beams propagated perpendicular to the homogeneous magnetic field, which defined the quantization axis throughout the experiment. The linear polarization of each laser was controlled independently using half-wave plates. Throughout this work, $P$ polarization denotes linear polarization parallel to the quantization axis and therefore drives transitions with $\Delta m=0$, whereas $S$ polarization denotes linear polarization perpendicular to the quantization axis. The latter can be decomposed into equal $\sigma^+$ and $\sigma^-$ components and consequently drives transitions with $\Delta m=\pm1$.

\subsection{COLTRIMS detection and event-by-event time reconstruction}
\label{sec:coltrims}

Charged reaction products were detected using a COLTRIMS reaction microscope \cite{Hubele2015}. Electrons and recoil ions were extracted by a homogeneous electric field of approximately $1\,\mathrm{V/cm}$ and guided by a homogeneous magnetic field of about $4\,\mathrm{G}$ onto opposing time- and position-sensitive microchannel-plate detectors equipped with delay-line anodes. The detector signals provide the particle impact positions and arrival times. When used with a pulsed light source, the laser pulse provides a temporal reference for the ionization event, allowing these observables to be used to reconstruct the three-dimensional electron and ion momenta. For the continuous-wave light fields employed here, however, no such external timing reference is available. The unknown ionization time therefore introduces a common offset in the particle times of flight and prevents a direct reconstruction of the longitudinal momentum components.

The ionization time was determined on an event-by-event basis using the reconstruction method developed in Ref.~\cite{Romans2025}. The method varies the common start time until the reconstructed longitudinal electron and recoil-ion momenta satisfy momentum conservation. The resulting ionization time then permits reconstruction of the complete three-dimensional momentum vectors for each coincidence event.

\section{RESULTS AND DISCUSSION}

\begin{figure}
    \centering
    \includegraphics[width=\linewidth]{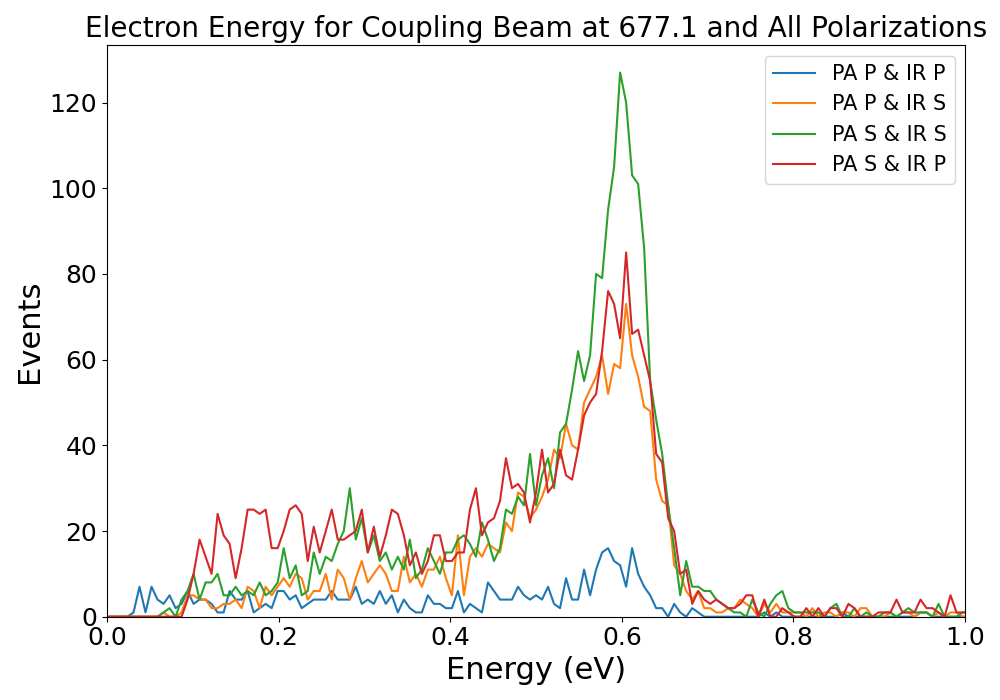}
\caption{\label{fig:evplot677}
Photoelectron energy spectrum obtained for the 677.1\,nm excitation scheme. The data include all four combinations of coupling and trapping laser polarizations.
}    
\end{figure} 

\begin{figure}
    \centering
    \includegraphics[width=\linewidth]{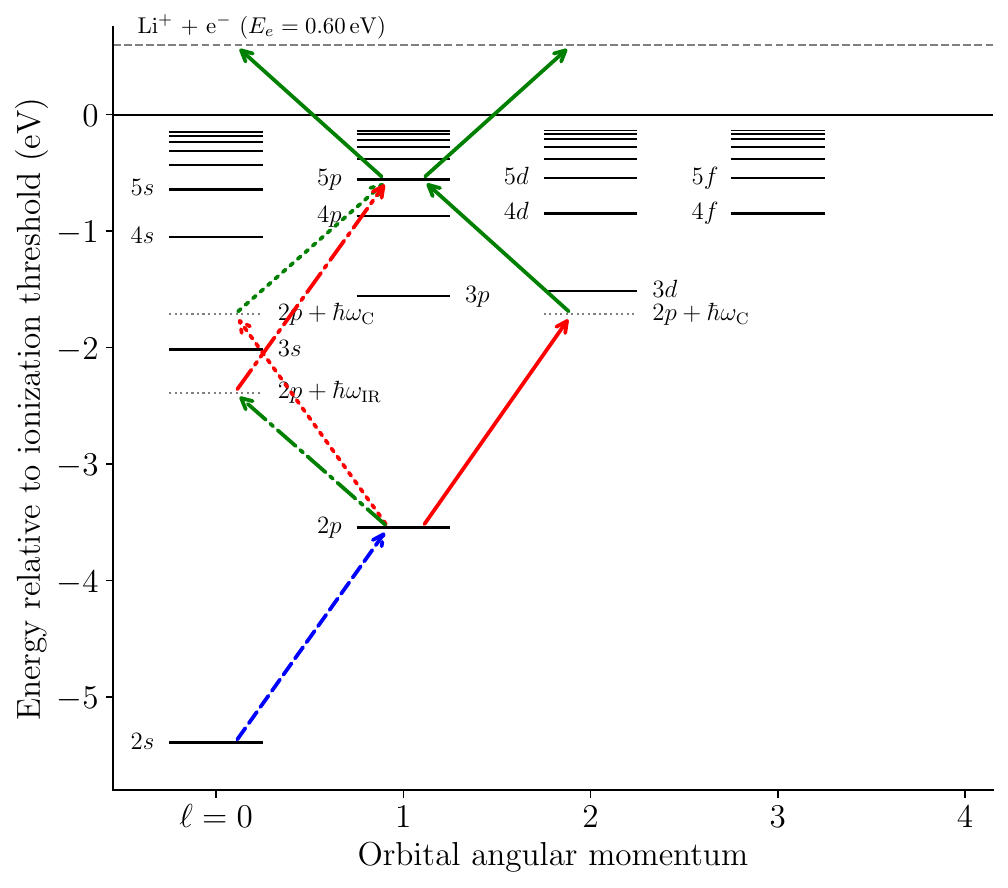}
    \caption{
    \label{fig:leveldiagram677}
    Simplified excitation scheme illustrating the resonant three-photon ionization process investigated in this work. Atoms are prepared in the $2p$ state by the cooling laser (blue), and are resonantly excited to the $5p$ state by absorption of one 677.1\,nm coupling photon (red) and one 1070\,nm infrared photon (green). Horizontal dashed lines indicate the energies of the virtual intermediate states reached after absorption of the first photon and do not represent stationary atomic levels. Solid, dotted, and dash-dotted arrows denote the dominant excitation through virtual states of $d$ symmetry and the two possible time orderings through virtual states of $s$ symmetry, respectively. The corresponding reverse time ordering via virtual $d$ states is omitted for clarity because of its much larger detuning from resonance. Absorption of an additional infrared photon ionizes the $5p$ state, producing photoelectrons with a kinetic energy of approximately 0.60\,eV, indicated by the upper dashed line.
    }
\end{figure}

At the laser intensities employed in this work, nonresonant three-photon ionization is expected to be strongly suppressed, and the observed photoelectron signal is therefore dominated by resonance-enhanced multiphoton ionization (REMPI) \cite{Delone2000}. We first consider measurements performed with the 677.1\,nm coupling laser in combination with the 1070\,nm trapping laser.

Figure~\ref{fig:evplot677} shows the measured photoelectron energy spectra for four combinations of coupling and infrared laser polarizations. All spectra are dominated by a single peak at a kinetic energy of approximately 0.60\,eV. This observation is consistent with resonant excitation of the $5p$ state through the absorption of one 677.1\,nm coupling photon and one 1070\,nm infrared photon, followed by absorption of an additional infrared photon that ionizes the atom. The corresponding excitation scheme is illustrated in Fig.~\ref{fig:leveldiagram677}. Although other ionization pathways are, in principle, possible, they are not resolved in the measured spectra. In particular, ionization by absorption of a second 677.1\,nm photon is expected to be less probable because the intensity of the coupling laser is substantially lower than that of the infrared trapping laser.

While all polarization combinations produce photoelectrons with the same kinetic energy, the photoelectron yield varies considerably. In the following, the notation ($P|S$) specifies the polarization of the coupling laser (first letter) and the infrared laser (second letter). The weakest signal is observed for the $P|P$ configuration, whereas the remaining three polarization combinations produce substantially higher count rates. Although the measurements do not provide absolute ionization probabilities, each polarization configuration was recorded under identical experimental conditions and for the same acquisition time. Throughout the measurement sequence, the relevant experimental parameters, including the trapped atom number and laser operating conditions, were continuously monitored and remained sufficiently stable that the relative ionization rates are estimated to be reproducible to better than 20\%. The observed differences between the polarization configurations therefore significantly exceed the experimental fluctuations.

This pronounced polarization dependence originates from the combination of the optically prepared initial state and the selection rules governing the resonant two-photon transition. As discussed in Secs.~\ref{sec:aot} and \ref{sec:lasers}, the lithium atoms are prepared predominantly in the $2p(m = +1)$ magnetic sublevel by optical pumping in the presence of the homogeneous magnetic field. The P- and S-polarized laser fields, defined in Sec.~\ref{sec:lasers}, drive transitions with $\Delta m=0$ and $\Delta m=\pm1$, respectively.

The resonant excitation from the $2p$ to the $5p$ state is a two-photon process and is therefore described by second-order perturbation theory. The corresponding transition amplitude is
\begin{equation}
\begin{aligned}
M^{(2)} = \sum_\nu \Biggl[ &
\frac{\langle 5p|\hat d_{\mathrm{IR}}|\nu\rangle
\langle \nu|\hat d_{\mathrm C}|2p\rangle}{E_{2p}+\hbar\omega_{\mathrm C}-E_\nu} \\
& +
\frac{
\langle 5p|\hat d_{\mathrm C}|\nu\rangle
\langle \nu|\hat d_{\mathrm{IR}}|2p\rangle
}{
E_{2p}+\hbar\omega_{\mathrm{IR}}-E_\nu
}
\Biggr],
\end{aligned}
\label{eq:secondorder}
\end{equation}
where the sum extends over all allowed intermediate states $|\nu\rangle$, with energies $E_\nu$. The operators $\hat d_{\mathrm C}=\hat{\mathbf d}\cdot\boldsymbol{\epsilon}{\mathrm C}$ and $\hat d{\mathrm{IR}}=\hat{\mathbf d}\cdot\boldsymbol{\epsilon}_{\mathrm{IR}}$ describe the atomic electric dipole operator projected onto the polarization directions of the coupling and trapping laser fields, respectively. The two terms represent the two possible time orderings of the absorbed photons, while the energy denominators describe the detuning between the intermediate-state energies and the energies reached after absorption of the first photon.

Figure~\ref{fig:leveldiagram677} illustrates excitation pathways contributing to the second-order transition amplitude in Eq.~(\ref{eq:secondorder}). Only intermediate states for which both dipole matrix elements are non-zero contribute to the coherent sum. The figure highlights excitation pathways via intermediate states of both $d$ and $s$ symmetry. For clarity, only the time ordering in which the coupling photon is absorbed first is shown for the pathway involving intermediate states of $d$ symmetry. The reverse time ordering is fully included in Eq.~(\ref{eq:secondorder}), but is omitted from the figure because the corresponding energy denominator is much larger, resulting in a substantially smaller contribution to the total transition amplitude.

We first examine the excitation pathways involving virtual states of $s$ symmetry. From the initial $2p(m=+1)$ state, a transition to a virtual $s$ state requires $\Delta m=-1$. Consequently, this pathway can only be driven by an $S$-polarized photon, whose $\sigma^-$ component provides the required change in magnetic quantum number. By contrast, a $P$-polarized photon ($\Delta m=0$) cannot couple the initial state to the virtual $s$ state. Thus, excitation pathways via virtual $s$ states contribute to the sum in Eq.~(\ref{eq:secondorder}) only when at least one of the absorbed photons is $S$ polarized. The additional excitation pathways available for polarization configurations containing an $S$-polarized photon provide a natural explanation for the experimentally observed increase in the photoionization yield relative to the $P|P$ configuration.




\begin{figure}
    \centering
    \includegraphics[width=\linewidth]{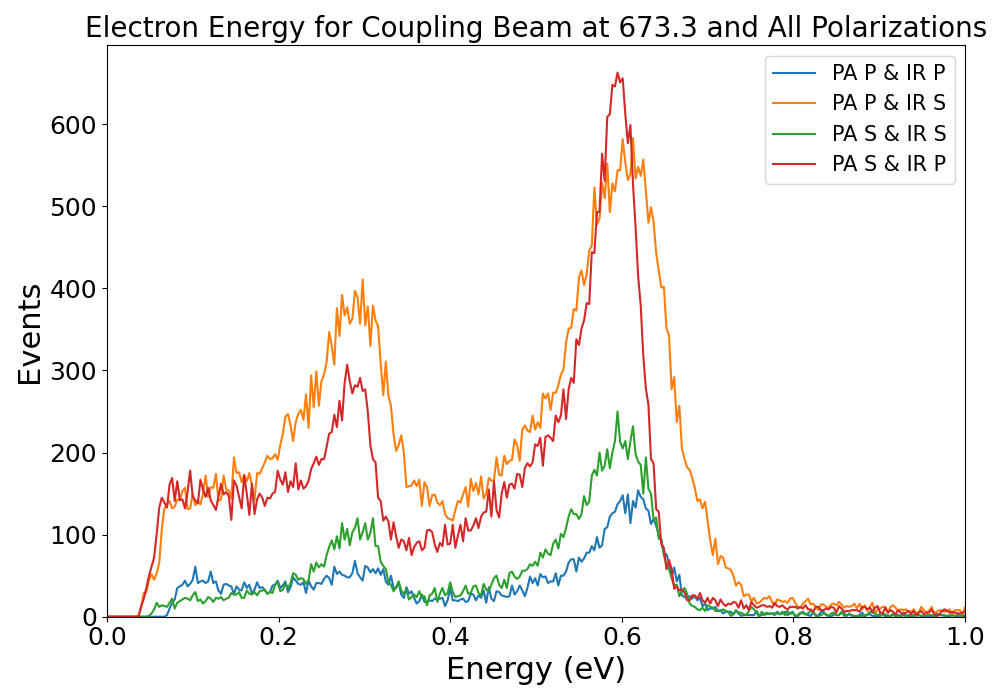}
    \caption{\label{fig:evplot673}Photoelectron energy spectrum corresponding to the 673.3\,nm excitation scheme. The data were obtained for all four combinations of coupling and trapping laser polarizations.}
\end{figure} 

\begin{figure}
    \centering
    \includegraphics[width=\linewidth]{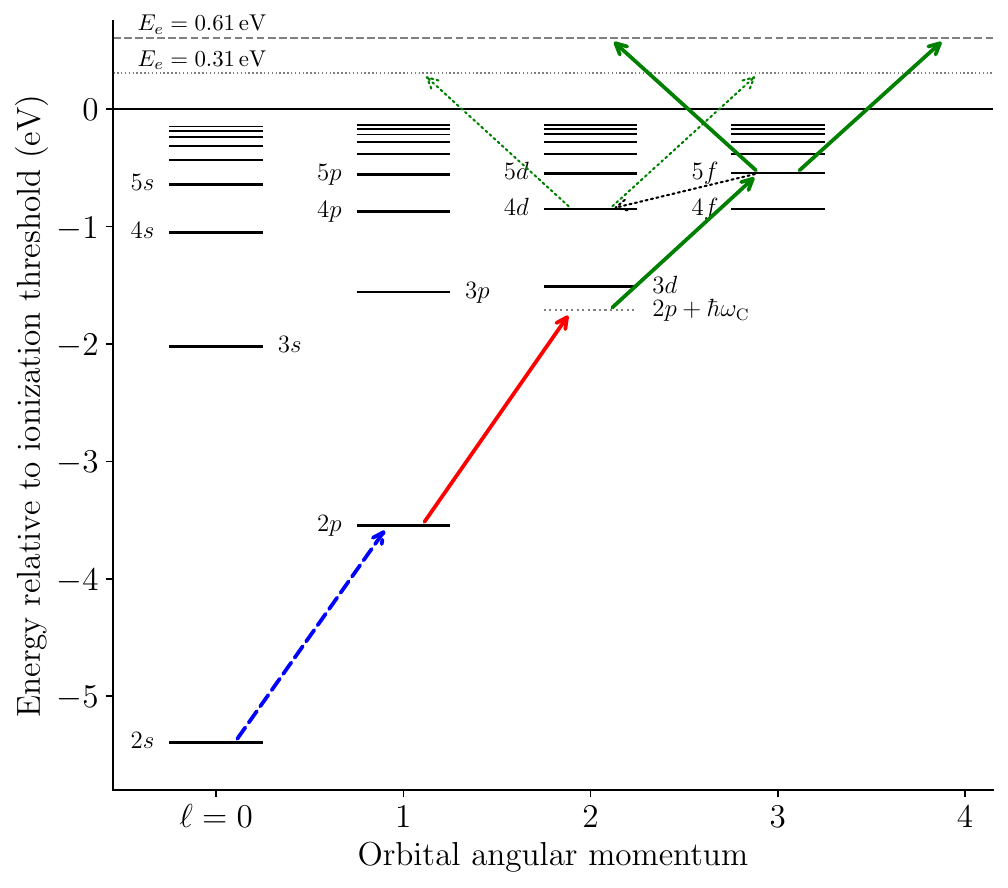}
    \caption{
    \label{fig:leveldiagram673}
    Same as Fig.~\ref{fig:leveldiagram677}, but for the 673.3\,nm coupling laser driving the resonant $2p \rightarrow 5f$ transition. The dotted arrows indicate the additional ionization pathway via spontaneous decay to the 4d state.
    }
\end{figure}

We now turn to measurements performed with the 673.3\,nm coupling laser in combination with the 1070\,nm trapping laser. Figure~\ref{fig:evplot673} shows the corresponding photoelectron energy spectra. Compared with the 677.1\,nm excitation scheme discussed above, two notable differences are observed. First, the overall photoelectron yield is substantially higher. Second, an additional photoelectron peak appears at a kinetic energy of approximately 0.31\,eV. The dominant peak at approximately 0.61\,eV is consistent with resonant excitation of the $5f$ state through absorption of one coupling photon and one infrared photon, followed by absorption of an additional infrared photon that ionizes the atom. The corresponding excitation scheme is illustrated in Fig.~\ref{fig:leveldiagram673}.

The enhanced photoelectron yield can be understood by comparing the dominant spontaneous decay channels and estimated photoionization rates for the resonantly excited $5p$ and $5f$ states, summarized in Table~\ref{tab:decay_ionization}. The most striking difference is that the spontaneous decay rate for the $5f\rightarrow3d$ transition is approximately twenty times larger than that for the $5p\rightarrow3d$ transition. Since spontaneous decay rates are proportional to the squared transition dipole matrix elements, the substantially larger decay rate indicates a stronger $3d-5f$ dipole coupling. This coupling also enters the second-order excitation amplitude in Eq.~(\ref{eq:secondorder}) through the virtual $3d$ intermediate state and can therefore enhance the population transferred to the 5f state. 

The photoionization rates listed in Table~\ref{tab:decay_ionization} were calculated using hydrogenic bound--free Gaunt factors within the Coulomb approximation \cite{Karzas1961,ZeQing1998} and the photon flux corresponding to a peak infrared intensity of approximately $1\,\mathrm{MW/cm^2}$. The quoted values are averaged over the allowed bound--continuum channels; the exact rates depend on the laser polarization and the populated magnetic sublevels. Interestingly, the estimated photoionization rate from the $5f$ state is approximately a factor of three smaller than that from the $5p$ state, which at first sight appears inconsistent with the substantially larger photoelectron yield observed experimentally. However, both photoionization rates remain much larger than the expected resonant excitation rate under the present experimental conditions. Photoionization is therefore not the rate-limiting step. Instead, the overall photoelectron yield is determined primarily by the population transferred to the resonantly excited state. The substantially stronger $3d\leftrightarrow5f$ transition can therefore lead to a larger $5f$ population, more than compensating for its somewhat smaller photoionization rate and resulting in the enhanced photoelectron yield.

\begin{table}[t]
\centering
\begin{tabular}{lccc}
\hline\hline
Process & $5p$ rate (s$^{-1}$) & $5f$ rate (s$^{-1}$) & $5f/5p$ \\
\hline
Spont.\ decay into $3d$
    & $2.1\times10^{5}$
    & $4.6\times10^{6}$
    & 21.9 \\
Spont.\ decay into $4d$
    & $2.5\times10^{5}$
    & $2.6\times10^{6}$
    & 10.4 \\
IR photoionization
    & $5.3\times10^{7}$
    & $1.7\times10^{7}$
    & 0.32 \\\hline
\end{tabular}
\caption{Comparison of the dominant spontaneous decay channels and the estimated direct photoionization rates at 1070\,nm for the resonantly excited $5p$ and $5f$ states. The spontaneous decay rates were obtained by combining the relevant Einstein $A$ coefficients listed in the NIST Atomic Spectra Database \cite{NIST_ASD}. The photoionization rates were estimated for a peak IR intensity of approximately $1~\mathrm{MW/cm^2}$.}
\label{tab:decay_ionization}
\end{table}

The second notable difference between the two excitation schemes is the appearance of the additional photoelectron peak at approximately 0.31\,eV. This kinetic energy identifies the ionization as originating from the $4d$ state, since absorption of one infrared photon from this level produces photoelectrons with the observed energy. The absence of this feature for excitation via the $5p$ state can again be understood from the competition between spontaneous decay and photoionization summarized in Table~\ref{tab:decay_ionization}. For the $5p$ state, photoionization is much faster than spontaneous decay into the $4d$ state, such that most atoms are ionized before they can decay. For the $5f$ state, by contrast, spontaneous decay competes much more effectively with direct photoionization, allowing a substantial fraction of the excited-state population to populate the $4d$ state before ionization. Subsequent absorption of one infrared photon then produces the additional low-energy photoelectron peak.

As for the 677.1\,nm excitation scheme, the total photoelectron yield also exhibits a pronounced polarization dependence, with the crossed-polarization configurations producing larger yields than the corresponding parallel-polarization configurations. This behavior indicates that the excitation and ionization probabilities depend on the relative polarizations of the two fields through the angular-momentum coupling of the contributing pathways. A quantitative interpretation would require a magnetic-sublevel-resolved calculation and is beyond the scope of the present work.

\begin{figure}
    \centering
    \includegraphics[width=\linewidth]{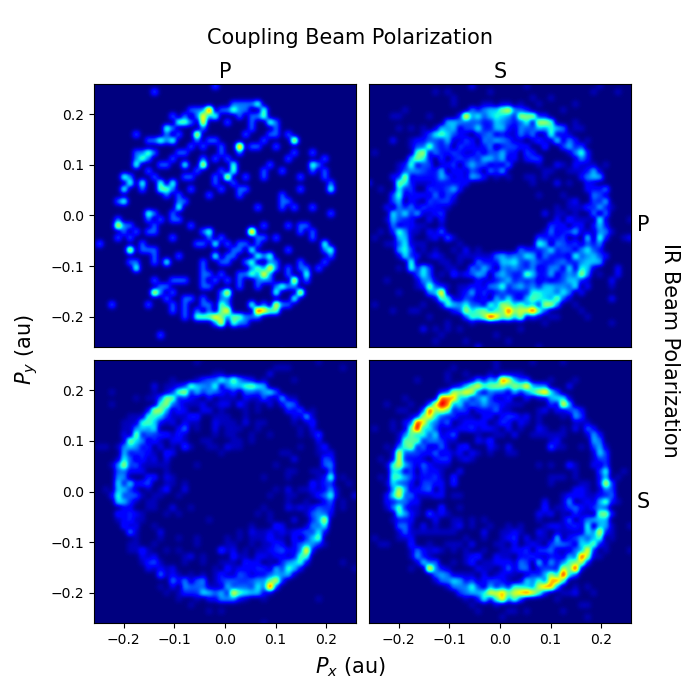}
    \caption{\label{fig:All_677.1_AllPXYe}Photoelectron momentum distributions in the $xy$ plane obtained for the 677.1\,nm excitation scheme. The four panels correspond to all combinations of coupling- and infrared-laser polarizations, with the coupling-laser polarization indicated above each column and the infrared-laser polarization indicated to the right of each row.}
\end{figure} 

We next examine the photoelectron momentum distributions obtained for the 677.1\,nm excitation scheme. Figure~\ref{fig:All_677.1_AllPXYe} shows the corresponding projection of the electron momentum onto the $xy$ plane for all four combinations of coupling- and trapping-laser polarizations. In each case, the dominant feature is a ring corresponding to the photoelectron peak near 0.60\,eV. The pronounced variation of the intensity around the ring demonstrates that the photoelectron angular distribution depends strongly on the relative polarizations of the two fields. The central region has been excluded because the momentum reconstruction provides insufficient resolution for electrons detected close to the spectrometer axis.

To analyze the observed momentum distributions, we follow the general approach used in our previous study of magnetic dichroism in few-photon ionization of polarized atoms \cite{Acharya2021}. We consider the azimuthal photoelectron angular distribution in the $xy$ plane perpendicular to the quantization axis ($z$). For photoelectrons with a fixed kinetic energy emitted into this plane, the azimuthal dependence of the continuum wave function can be expanded as
\begin{equation}
\Psi(\varphi)=\sum_m c_m e^{im\varphi},
\label{eq:phiexpansion}
\end{equation}
where the complex coefficients $c_m$ describe the amplitudes of the contributing continuum magnetic sublevels. The corresponding photoelectron angular distribution is then given by
\begin{equation}
\begin{aligned}
\frac{\mathrm{d}\sigma}{\mathrm{d}\Omega}
&=
\left|\Psi(\varphi)\right|^2 \\
&=
\sum_m |c_m|^2
+
\sum_{m<m'} 2|c_m||c_{m'}| \\
&\qquad\times
\cos\!\left[(m'-m)\varphi+\Delta_{m'm}\right],
\end{aligned}
\label{eq:phidistribution}
\end{equation}
where $\Delta_{m'm}=\arg(c_{m'})-\arg(c_m)$ denotes the relative phase between the interfering partial-wave amplitudes. The azimuthal angular distribution therefore consists of a constant background and a superposition of cosine modulations. Their amplitudes are determined by the magnitudes of the contributing magnetic sublevels, whereas their angular orientation is governed by the corresponding relative phases. These phases depend on the polarization directions of the excitation fields as well as on the dynamical phases accumulated during the multi-photon excitation and ionization process.

The qualitative features of the measured angular distributions can be understood from the magnetic-sublevel pathways allowed by the different polarization combinations. For the $P|P$ configuration, the magnetic quantum number is conserved throughout the excitation and ionization sequence, such that only the $m=+1$ continuum component is populated. Since only a single magnetic sublevel contributes, no interference is possible and the angular distribution is expected to be isotropic in the azimuthal direction. The measured momentum distribution is broadly consistent with this expectation within the statistical uncertainty of the data.

For an $S$-polarized coupling field and a $P$-polarized infrared field, the resonant excitation populates the $5p$  state with $m=0$. Because the second infrared photon also preserves the magnetic quantum number, only the $m=0$ continuum component is populated and  no azimuthal interference is expected. Although the measured momentum distribution exhibits some intensity variations, no clear modulation pattern is observed.


For a $P$-polarized coupling field and an $S$-polarized infrared field, the resonant excitation populates the $5p$, $m=0$ state. Absorption of the second infrared photon subsequently produces continuum components with $m=-1$ and $m=+1$. Their coherent interference gives rise to a twofold azimuthal modulation whose orientation is fixed by the infrared polarization axis. 

Finally, when both fields are $S$ polarized, continuum components with $m=-2$, $0$, and $+2$ may be populated. Interference between these components gives rise to the observed azimuthal modulation. In contrast to the $P|S$ configuration, the interference pattern is not required by symmetry to be aligned with the infrared polarization axis, and an azimuthal rotation associated with magnetic dichroism is, in principle, possible. Experimentally, however, the measured distribution is dominated by a twofold modulation, characterized by two maxima and two minima over a full rotation, and no significant angular rotation is observed within the experimental uncertainty.

\begin{figure}[t]
    \centering
    \includegraphics[width=\linewidth]{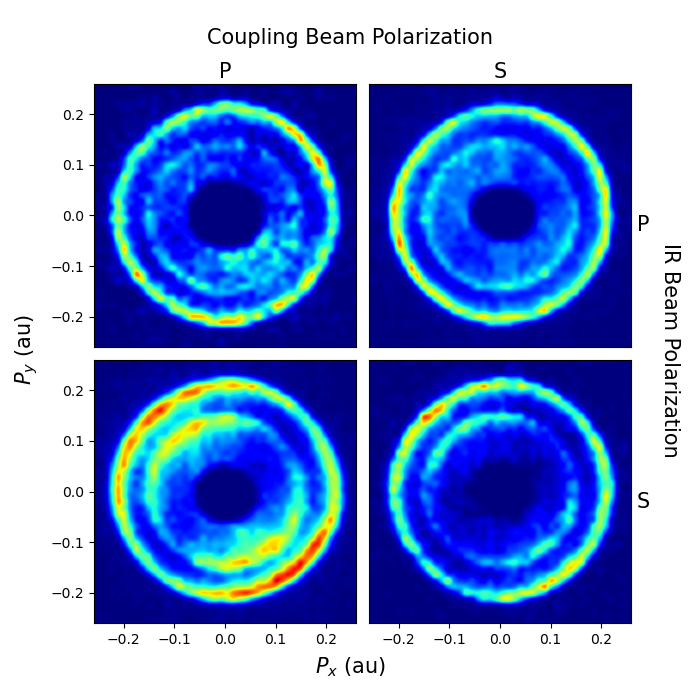}
    \caption{\label{fig:All_673.3_AllPXYe}Same as Fig.~\ref{fig:All_677.1_AllPXYe}, but for the 673.3\,nm excitation scheme.}
\end{figure} 

The momentum distributions obtained for the 673.3\,nm excitation scheme exhibit a similar qualitative polarization dependence (see Fig.~\ref{fig:All_673.3_AllPXYe}). For the direct ionization channel from the $5f$ state, the allowed continuum magnetic sublevels are $m=+1$ for $P|P$, $m=0,+2$ for $S|P$, $m=-1,+1,+3$ for $P|S$, and $m=-2,0,+2,+4$ for $S|S$. No clear azimuthal modulation is observed for the first two configurations. Although both the $m=0$ and $m=+2$ continuum magnetic sublevels are allowed for the $S|P$ configuration, the measured momentum distribution does not exhibit a pronounced modulation. This may indicate that the corresponding continuum amplitudes are strongly imbalanced, resulting in only weak interference.

The $P|S$ and $S|S$ distributions exhibit pronounced two-lobe structures. This behavior is qualitatively consistent with interference between the continuum magnetic sublevels, with the observed distributions predominantly arising from interference terms with $\Delta m=2$. Interestingly, the lower-energy photoelectron peak arising from spontaneous decay into the $4d$ state followed by ionization exhibits the same qualitative polarization dependence. Although spontaneous emission may redistribute the magnetic-sublevel population and reduce coherences, this observation suggests that polarization-dependent alignment is at least partially retained through the decay process. A quantitative description would, however, require an explicit treatment of the spontaneous-emission and ionization dynamics.

Taken together, the measured energy spectra, photoelectron yields, and momentum distributions consistently support the resonant excitation pathways proposed for the 677.1\,nm and 673.3\,nm excitation schemes. The observed photoelectron yields are qualitatively explained by the resonant transition strengths and the competing excitation, decay, and ionization dynamics, while the azimuthal angular distributions follow naturally from the magnetic-sublevel selection rules and their coherent interference.


\section{CONCLUSION}
We have investigated continuous-wave multicolor photoionization of laser-cooled lithium atoms confined in an all-optical trap using momentum-resolved reaction microscopy. By applying an event-by-event photoionization-time retrieval technique, complete three-dimensional electron momentum distributions and kinetic-energy spectra were obtained despite the absence of a pulsed ionization source.

For coupling-laser wavelengths of 677.1\,nm and 673.3\,nm, the measured electron spectra identify resonant two-photon excitation from the $2p$ state to the $5p$ and $5f$ states, respectively, followed by ionization through absorption of an additional infrared photon. Both excitation schemes produce a dominant photoelectron peak near 0.60\,eV. The ionization yield obtained through the $5f$ resonance is substantially larger than that observed through the $5p$ resonance. In addition, excitation of the $5f$ state produces a second, lower-energy photoelectron channel, which is attributed to spontaneous decay to the $4d$ state followed by infrared ionization.

The measured yields and momentum distributions exhibit a pronounced dependence on the polarization combination of the coupling and infrared laser fields. For the $5p$ resonance, the particularly weak yield observed for the $P|P$ configuration is consistent with the reduced number of allowed virtual excitation pathways imposed by the magnetic-sublevel selection rules. More generally, the polarization dependence demonstrates the sensitivity of the photoelectron signal to the coherent contributions of different excitation and ionization pathways.

These results show that momentum-resolved photoelectron spectroscopy can provide direct access to the microscopic excitation, decay, and ionization processes occurring during the operation of an optical atom trap. The approach therefore complements conventional measurements of atom loss and total ion yield and offers a route toward detailed studies of weak-field, continuous-wave light--matter interactions in laser-cooled atomic systems.

\section{ACKNOWLEDGEMENTS}
This work was supported by the U.S.\ National Science Foundation under Grant No.\ PHY-2207854.

\end{document}